\RequirePackage{fix-cm}
\documentclass[12pt]{article}                     
\usepackage{graphicx}
\usepackage{hyperref}
\usepackage{amsmath}
\usepackage{amsthm}
\usepackage{amsfonts}
\usepackage{hyperref}
\usepackage[top=1in, bottom=1in, left=1in, right=1in]{geometry}
\begin{document}

\title{Mean field limit for many-particle interactions
}


\author{Can Gokler\footnote{Harvard University, Cambridge MA, USA}}




\date{\vspace{-6ex}}

\maketitle

\abstract
We provide an error bound for approximating the time evolution of N bosons by a generalized nonlinear Hartree equation. The bosons are  assumed to interact via permutation symmetric bounded many-particle potentials and the initial wave-function is a product state. We show that the error between the actual evolution of a single particle derived from tracing out the full N-particle Schrodinger equation and the solution to the mean field approximate generalized nonlinear Hartree equation scales as 1/N for all times. Our result is a generalization of rigorous error bounds previously given for the case of bounded 2-particle potentials.

\section{Introduction}

\indent

Consider N identical bosons described on the permutation symmetric subspace of the Hilbert space $\mathcal{H}_1 \otimes ... \otimes \mathcal{H}_N$, $\dim \mathcal{H}_j = d$, where the Hamiltonian governing their time evolution is invariant under permutations of particles. The interaction potential on m-particles is a Hermitian operator denoted by $V^{(m)}_{j_1...j_m} $ which is identity on every particle except particles $j_1, ..., j_m$. Let the Hamiltonian consist of 1 up to M particle interactions:

\begin{equation}
H_N = \sum_{j=1}^N V^{(1)}_j + \frac{1}{N} \sum_{1=i <  j}^N V^{(2)}_{ij} +  \frac{1}{N^2} \sum_{1=i < j < k}^N V^{(3)}_{ijk}+...+ \frac{1}{N^{M-1}} \sum_{1=j_1< ... < j_M}^N V^{(M)}_{j_1 ... j_M}
\end{equation}

The $\frac{1}{N^j}$ factors are included so that the total strength of $k$ and $l$-particle interactions are of the same order in N. Let $\text{tr}_{[j,N]}$ is the partial trace over $j, j+1, ..., N$th particles. 

If initially the bosons are in a product state $\gamma_N (0)= \gamma(0)^{\otimes N} $ where $\gamma (0)= | \phi(0) \rangle \langle \phi(0)|$, the many-particle evolution can be approximated by a tensor product evolution governed by the generalized nonlinear Hartree equation:

\begin{equation}
i \frac{d}{dt} \gamma(t) = [V^{(1)}, \gamma(t)] + \sum_{j=2}^M \text{tr}_{[2, j]} [V^{(j)}, \gamma(t)^{\otimes j}]
\end{equation}

The widely known Hartree equation corresponds to the case $M=2$, where only two body potentials are considered. The first rigorous results proving the convergence of the solution of the Hartree equation to the actual evolution of 1-particle density matrix as  $N \rightarrow  \infty$ appeared in \cite{Hepp, Spohn, Spohn2}. The rate of convergence towards Hartree dynamics was investigated in \cite{Rodnianski, ChenSchlein, Knowles, Pickl, Ammari, Schlein}. The case $M=3$ was studied in \cite{ChenPavlovic, Chen}. See \cite{Schleinbook} for a review of rigorous mean field limits in quantum systems. In this paper we consider the case of general $M$ and give an error bound closely following the methods of Erd\H{o}s-Schlein\cite{Schlein}. Our main result is the following bound for the approximation error:

\indent

\noindent
\textbf{Theorem 1.}
Let $\gamma(t) = | \phi (t)\rangle \langle \phi (t) | $ be the solution of the generalized nonlinear Hartree equation with product initial state $\gamma_N (0)= \gamma(0)^{\otimes N} $ where $\gamma (0)= | \phi(0) \rangle \langle \phi(0)|$. Let $\text{tr}_{[2, N]} (e^{-i H_N t} \gamma_N(0) e^{i H_N t})$ be the actual time evolution of the 1-particle density matrix. Then the following bound holds
\begin{equation}
\text{tr}|\gamma(t) - \text{tr}_{[2, N]} (e^{-i H_N t} \gamma_N(0) e^{i H_N t})| \leq \frac{M^3}{N} \lambda_V  (e^{4 (\sum l \cdot V) t} - 1 )
\end{equation}
where we made the following definitions

 \begin{equation}
 (\sum l^k \cdot V) = \sum_{l=2}^M l^k |V^{(l)}|
 \end{equation}
 
 \begin{equation}
 |\tilde{V}|= \text{max}_{\text{basis}}\text{max}_m \sum_{j_1...j_m} |w^{(m)}_{j_1...j_m}|
 \end{equation} 
 
 \begin{equation}
 \lambda_V=\frac{ 16|\tilde{V}|+ (\sum l^2 \cdot V)}{(\sum l \cdot V)} )
 \end{equation}. 
  
The relation of $w^{(m)}_{j_1...j_m} $ to $V^{(m)}$ and maximization over the bases are defined in the following way. Choose an orthonormal basis for operators acting on $m$-particles ${1,...,m}$ as $\{E_{1,i_1} \otimes... \otimes E_{m,i_m} \}$ with $\text{tr} E_{j,k_1}E_{j,k_2}^{\dagger}=\delta_{k_1,k_2}$ for all $j$ and decompose $V^{(m)}$ as $V^{(m)}=\sum_{i_1...i_m} w_{i_1...i_m}E_{1, i_1} \otimes... \otimes E_{m, i_m}$. Maximization is over all such orthonormal bases. We set $\hbar=1$ throughout the paper.

The proof has two main steps. First we provide Lieb-Robinson type bounds for the correlations between two observables acting on distinct groups of particles. In the second part of the proof we will utilize these bounds to truncate the BBGKY hierarchy to derive the main result. The mean field approximation neglects correlations between particles. The Lieb-Robinson type bounds give a way to estimate how much correlation one neglects. Inputting them to BBGKY hierarchy gives the error involved in approximating the dynamics by the mean field equation.

\section{Lieb-Robinson type bound on the growth of correlations between two observables acting on distinct groups of particles }

\indent

Our first step is to prove correlation bounds on two arbitrary observables acting on distinct sets of particles.  

\indent

\noindent
\textbf{Proposition 1. Lieb-Robinson bound for correlations.} Let $A$ and $B$ two bounded observables acting on distinct set of particles $ \mathcal{I}_m=\{ i_1, i_2, ..., i_m \}$ and $ \mathcal{J}_n=\{ j_1, j_2, ..., j_n \}$, respectively. Then

\begin{align}
 | [A, e^{iH_Nt} B e^{-iH_Nt}] | \leq \frac{4mn |A| |B|}{N} (e^{2 ( \sum l \cdot V)t} -1) 
\end{align}

Compare this with the generic form of the Lieb-Robinson bound on a lattice: 

\begin{equation}
| [A, e^{iH_Nt} B e^{-iH_Nt}] | < a e^{-b(d(A,B)-vt )}
\end{equation}
where $a$ and $b$ are constants, $A$ and $B$ are assumed to have distance $d(A,B)$ and $v$ is the effective velocity \cite{Lieb}.

\noindent
\textbf{Proof of Proposition 1.} First define the modified Hamiltonian $H_N^{(n)}$ which decouples $n$ particles from the rest as

\begin{equation}
H_N^{(n)} = H_N - \frac{1}{N} \sum_{j_1=1}^n \sum_{n<j_2}^N V^{(2)}_{j_1j_2} - \frac{1}{N^2} \sum_{j_1=1}^n \sum_{n<j_2<j_3}^N V^{(3)}_{j_1j_2j_3} - ... - \frac{1}{N^{M-1}}  \sum_{j_1=1}^n \sum_{n<j_2<...<j_M}^N V^{(M)}_{j_1...j_M}
\end{equation}

Note that the terms subtracted from the original Hamiltonian are exactly the ones which couple the first $n$ particles to the rest. Next define the normalized correlation $f_{mn}(t)$ as

\begin{equation}
f_{mn}(t)=\text{sup}_{A,B} \frac{| [A, e^{iH_Nt} B e^{-iH_Nt}] |}{|A| |B|}
\end{equation}

 $f_{mn}(t)$ does not depend on particular choices of $\mathcal{I}_m$ and $\mathcal{J}_n$ because of the permutation symmetry of $H_N$. Then take $\mathcal{I}_m=\{ n+1, ..., n+m \} $ and $ \mathcal{J}_n= \{ 1, ..., n \} $. Since $H_N^{(n)}$ acts only on first $n$ particles
 
 \begin{equation}
 f_{mn}(t)=\text{sup}_{A,B} \frac{| [A, e^{iH_Nt} e^{-iH_N^{(n)}t} B  e^{iH_N^{(n)}t} e^{-iH_Nt}] |}{|A| |B|}
 \end{equation}
 
 Then further define 
\begin{equation}
g_{AB}(t)= [A, e^{iH_Nt} e^{-iH_N^{(n)}t} B  e^{iH_N^{(n)}t} e^{-iH_Nt}]
\end{equation} 

We will derive the evolution of $g_{AB}(t)$ and bound $|g_{AB}(t)|$ using a Dyson series approach. First calculate $\dot{g}_{AB}(t)$:

\begin{equation*}
\dot{g}_{AB}(t) =i [A, [e^{iH_Nt} (H_N-H^{(n)}_N) e^{-iH_Nt}, e^{iH_Nt} e^{-iH_N^{(n)}t} B  e^{iH_N^{(n)}t} e^{-iH_Nt} ]] 
\end{equation*}

\begin{align}
=i[e^{iH_Nt} (H_N-H^{(n)}_N) e^{-iH_Nt}, g_{AB}(t)]- \nonumber
\\ 
i[e^{iH_Nt} e^{-iH_N^{(n)}t} B  e^{iH_N^{(n)}t} e^{-iH_Nt}, [A, e^{iH_Nt} (H_N-H^{(n)}_N) e^{-iH_Nt}]]
\end{align}
where last step follows from the Jacobi identity. Define the two parameter group of transformations $\mathcal{U}^{(n)}(t,s)$  such that $\mathcal{U}^{(n)}(t,t)=\text{Id}$ and 

\begin{equation}
i \partial_t \mathcal{U}^{(n)}(t,s) = e^{iH_Nt} (H_N-H^{(n)}_N) e^{-iH_Nt} \mathcal{U}^{(n)}(t,s) 
\end{equation}

Then 

\begin{multline}
\partial_t  (\mathcal{U}^{(n)}(0,t) g_{AB}(t) \mathcal{U}^{(n)}(t,0)) = \\ -i \mathcal{U}^{(n)}(0,t) [e^{iH_Nt} e^{-iH_N^{(n)}t} B  e^{iH_N^{(n)}t} e^{-iH_Nt}, [A,  e^{iH_Nt} (H_N-H^{(n)}_N) e^{-iH_Nt}]] \mathcal{U}^{(n)}(t,0)
\end{multline}

Initially, at time $t=0$, since $A$ and $B$ act on distinct sets of particles $g_{AB}=0$. Integrating the previous equation one has

\begin{multline}
g_{AB}(t)= \\ -i \int_0^t ds \mathcal{U}^{(n)}(t,s) [e^{iH_Ns} e^{-iH_N^{(n)}s} B  e^{iH_N^{(n)}s} e^{-iH_Ns},
  [A,  e^{iH_Ns} (H_N-H^{(n)}_N) e^{-iH_Ns}]] \mathcal{U}^{(n)}(s,t)
\end{multline}

Insert  $H_N^{(n)} - H_N$ to get

\begin{align}
g_{AB}(t)=-i \sum_{k=2}^M \frac{1}{N^{k-1}} \sum_{j_1=1}^n \sum_{n< j_2< ...<j_k}^N\int_0^t ds  \mathcal{U}^{(n)}(t,s) [e^{iH_Ns} e^{-iH_N^{(n)}s} B  e^{iH_N^{(n)}s} e^{-iH_Ns}, \nonumber
\\ [A,  e^{iH_Ns}  V^{(k)}_{j_1...j_k} e^{-iH_Ns}]] \mathcal{U}^{(n)}(s,t)
\end{align}

Now take the operator norm of both sides and apply $|U B U^{\dagger}|=|B|$ where $U$ is a unitary and $| [B,C]| \leq 2 |B| |C|$ which follows from the Cauchy-Schwartz inequality.

\begin{align}
& |g_{AB}(t)| \leq  \sum_{k=2}^M \frac{1}{N^{k-1}} \sum_{j_1=1}^n \sum_{n< j_2< ...<j_k}^N\int_0^t ds  |\mathcal{U}^{(n)}(t,s) [e^{iH_Ns} e^{-iH_N^{(n)}s} B  e^{iH_N^{(n)}s} e^{-iH_Ns}, \nonumber & \\
 &  [A,  e^{iH_Ns}  V^{(k)}_{j_1...j_k} e^{-iH_Ns}]] \mathcal{U}^{(n)}(s,t)|  \nonumber  \\
 & \leq  \sum_{k=2}^M \frac{1}{N^{k-1}} \sum_{j_1=1}^n \sum_{n< j_2< ...<j_k}^N\int_0^t ds  | [e^{iH_Ns} e^{-iH_N^{(n)}s} B  e^{iH_N^{(n)}s} e^{-iH_Ns}, [A,  e^{iH_Ns}  V^{(k)}_{j_1...j_k} e^{-iH_Ns}]]|  \nonumber  & 
 \\
& \leq  \sum_{k=2}^M \frac{2}{N^{k-1}} \sum_{j_1=1}^n \sum_{n< j_2< ...<j_k}^N\int_0^t ds  | e^{iH_Ns} e^{-iH_N^{(n)}s} B  e^{iH_N^{(n)}s} e^{-iH_Ns}| | [A,  e^{iH_Ns}  V^{(k)}_{j_1...j_k} e^{-iH_Ns}] |  \nonumber & \\
& =  \sum_{k=2}^M \frac{2 |B|}{N^{k-1}} \sum_{j_1=1}^n \sum_{n< j_2< ...<j_k}^N\int_0^t ds | [A,  e^{iH_Ns}  V^{(k)}_{j_1...j_k} e^{-iH_Ns}] | &
\end{align}

Lets decompose the sum $\sum_{j_1=1}^n \sum_{n< j_2< ...<j_k}^N$ into two sums one of which is

\begin{equation}
\delta=\sum_{j_1=1}^n \sum_{n+m< j_2< ...<j_k}^N
\end{equation}
 and call the rest $\sigma$. We will use this decomposition and use permutation symmetry in the following. The original sum has $n {N-n \choose k-1}$ terms while $\delta$ has $n {N-n-m \choose k-1}$ and $\sigma$ has $n {N-n \choose k-1} - n {N-n-m \choose k-1}$ terms. It can be shown that the number of terms in $\sigma$ is bounded by $2 n mN^{k-2}$ while the number of terms in the original sum is bounded by $n N^{k-1}$. Then below we apply these to bound $|g_{AB}(t)|$. We also utilize the permutation symmetry.

\begin{align}
& |g_{AB}(t)| \leq  \sum_{k=2}^M \frac{2 |B|}{N^{k-1}} (\sigma + \delta)\int_0^t ds | [A,  e^{iH_Ns}  V^{(k)}_{j_1...j_k} e^{-iH_Ns}] |  \nonumber  & \\ & \leq | \sum_{k=2}^M \frac{2 |B|}{N^{k-1}} \sigma 2 |A| |V^{(k)}| t + \sum_{k=2}^M \frac{2 |B|}{N^{k-1}} \delta \int_0^t ds | [A,  e^{iH_Ns}  V^{(k)}_{j_1...j_k} e^{-iH_Ns}] |  \nonumber 
& \\ & \leq \frac{8mn |A||B|}{N} t \sum_{k=2}^M |V^{(k)}| + \sum_{k=2}^M \frac{2 |B|}{N^{k-1}} \delta \int_0^t ds | [A,  e^{iH_Ns}  V^{(k)}_{j_1...j_k} e^{-iH_Ns}] |  \nonumber 
& \\ & \leq \frac{8mn |A||B|}{N} t \sum_{k=2}^M |V^{(k)}| + \sum_{k=2}^M 2n |B| \int_0^t ds | [A,  e^{iH_Ns}  V^{(k)}_{1, n+m+1, n+m+2,..., n+m+k-1} e^{-iH_Ns}] |  \nonumber  
&  \\ & = \frac{8mn |A||B|}{N} t \sum_{k=2}^M |V^{(k)}| + \sum_{k=2}^M 2n |A| |B| |V^{(k)}| \int_0^t ds \frac{| [A,  e^{iH_Ns}  V^{(k)}_{1, n+m+1, n+m+2,..., n+m+k-1} e^{-iH_Ns}] |}{|A| |V^{(k)}|}  \nonumber  
&  \\ \nonumber & \leq \frac{8mn |A||B|}{N} t \sum_{k=2}^M |V^{(k)}| + 2n |A| |B| \sum_{k=2}^M  |V^{(k)}| \int_0^t ds f_{mk}(s) \\
\end{align}

Then

\begin{align}
& f_{mn}(t)= \text{sup}_{A,B} \frac{|g_{AB}(t)|}{|A| |B|} \leq \frac{8mn}{N} t \sum_{k=2}^M |V^{(k)}| + 2n \sum_{k=2}^M  |V^{(k)}| \int_0^t ds f_{mk}(s) \nonumber & \\
&= (\sum V) (\frac{8mn}{N} t + 2n  \int_0^t ds f_{mk}(s)) &
\end{align}

This is the anticipated Dyson series like integral inequality. Now, recursively apply the inequality q times to get

\begin{align}
&f_{mn}(t) \leq \frac{8mn}{N} t (\sum V) + \frac{8mn}{N}  (\sum V) \sum_{r=1}^{q-1} 2^r (\sum l \cdot V)^r \int_0^t ds_1 \int_0^{s_1} ds_2 ... \int_0^{s_r-1}ds_r s_r \nonumber &
\\ & + 2n 2^{q-1} (\sum l \cdot V)^{q-1} \sum_{k=2}^M |V^{(k)}| \int_0^t ds_1 \int_0^{s_1}ds_2 ... \int_0^{s_{q-1}} ds_q f_{mk}(s_q)&
\end{align}

Using the trivial bound $f_{mk} \leq 2$, we obtain

\begin{align}
&f_{mn}(t) \leq (4mn \sum_{r=0}^{q-1} \frac{(2(\sum l \cdot V) )^{r+1}}{(r+1)!} )+ 4n 2^q (\sum l \cdot V)^{q-1} (\sum V) \frac{t^q}{q!} &
\end{align}

Letting q to infinity we get

\begin{align}
f_{mn}(t) \leq \frac{4mn}{N} (e^{2 (\sum l \cdot V) t} - 1)
\end{align}

This gives the correlation bound

\begin{align}
 | [A, e^{iH_Nt} B e^{-iH_Nt}] | \leq \frac{4mn |A| |B|}{N} (e^{2 ( \sum l \cdot V)t} -1) 
\end{align}

\noindent
\textbf{Q.E.D.}

\indent

Following Erd\H{o}s-Schlein\cite{Schlein}, more specific correlation bounds are obtained. We have the two following bounds as corollaries of Proposition 1:

\indent

\noindent
\textbf{Corollary 1.} Let $A$ and $B$ two bounded observables acting on distinct set of particles $ \mathcal{I}_m=\{ i_1, i_2, ..., i_m \}$ and $ \mathcal{J}_n=\{ j_1, j_2, ..., j_n \}$, respectively. Let the N particle system be initially in the pure state $\psi_N(0)= | \phi \rangle^{ \otimes N}$ and let $\psi_N(t)=e^{-iH_Nt}\psi_N(0)$. Then
\begin{align}
| \langle \psi_N(t), AB \psi_N(t) \rangle - \langle \psi_N(t), A\psi_N(t) \rangle \langle \psi_N(t),B\psi_N(t) \rangle | \leq \frac{16mn |A| |B|}{N} (e^{4 ( \sum l \cdot V)t} -1)
\end{align}
and

\noindent
\textbf{Corollary 2.} Let $A$ and $B$ two bounded observables acting on distinct set of particles $ \mathcal{I}_m=\{ i_1, i_2, ..., i_m \}$ and $ \mathcal{J}_n=\{ j_1, j_2, ..., j_n \}$, respectively. Let the N particle system be initially in the pure state $\psi_N(0)= | \phi \rangle^{ \otimes N}$ and let $\psi_N(t)=e^{-iH_Nt}\psi_N(0)$. Then

\begin{align} \label{eq:1}
| \text{tr} (A \otimes B ) ( \gamma_N^{(m+n)}(t)-\gamma_N^{(m)}(t) \otimes \gamma_N^{(n)}(t) )| \leq \frac{16mn |A| |B|}{N} (e^{4 ( \sum l \cdot V)t} -1)
\end{align}
where $ \gamma_N^{(m)}(t)= \text{tr}_{[m+1, N]} | \psi_N(t) \rangle \langle \psi_N(t) |$.

\section{BBGKY hierarchy and the rate of convergence towards the Hartree limit}

\indent

BBGKY (Bogoliubov, Born, Green, Kirkwood, Yvon) hierarchy describes $k$-particle density evolution in terms of $m>k$ particle density matrices. The hierarchy equation for the $k$-particle density is obtained by tracing out the remaining particles in the Liouville-von Neumann equation $\gamma_N(t)=-i[H_N, \gamma_N(t)]$ for all $N$ particles. The set of hierarchy equations for all $k=1,...,N$ are an exact reformulation of the Liouville-von Neumann equation. In the following we want to truncate the hierarchy equations comparing them to the nonlinear Hartree evolution via trace distance using the correlation bounds given above in Proposition 1 and its corollaries. 

\indent

Consider only the $m$-body interaction term $W^{(m)}=\frac{1}{N^{m-1}} \sum_{1 \leq j_1 < ... < j_m}^N V_{j_1...j_m}^{(m)}$. The BBGKY hierarchy equation for this $m$-particle potential is given by (which is a generalization of the hierarchy given in \cite{Spohn})

\begin{align}
i \dot{\gamma}_N^{(k)}(t) =  \sum_{l=0}^{m-1} \frac{(N-k)!}{(N-k-l)!} \frac{1}{N^{m-1}} \sum_{1 \leq j_1 <...<j_{m-l}}^{k} \text{tr}_{[k+1, ..., k+l]} [ V^{(m)}_{j_1...j_{m-l}, k+1, ..., k+l}, \gamma_N^{(k+l)}(t)]
\end{align}

\noindent
where $k \geq m$. If all particle interactions are included then we need to sum over all $m$ on the right hand side. The validity of the BBGKY hierarchy equation can be shown noting the trace of Liouville-von Neumann equation:

\begin{equation}
i\dot{\gamma}_N^{(k)}(t)= i \text{tr}_{[k+1, N]}\dot{\gamma}_N(t)=\text{tr}_{[k+1, N]}[H_N^{(m)},\gamma_N(t)]
\end{equation}
and decomposing $W^{(m)}$ as below and using permutation symmetry upon taking the trace.

\begin{align}
N^{m-1}W^{(m)} =  \sum_{1 \leq j_1 < ... < j_m}^k V_{j_1...j_m}^{(m)} + \sum_{k < j_1 < ... < j_m}^N V_{j_1...j_m}^{(m)} + \sum_{1 \leq j_1 < ... < j_{m-1}}^k \sum_{k < j_m}^N V_{j_1...j_m}^{(m)} \nonumber
\\ + \sum_{1 \leq j_1 < ... < j_{m-2}}^k \sum_{k < j_{m-1} < j_m}^N V_{j_1...j_m}^{(m)} + ... + \sum_{1 \leq j_1}^k \sum_{k < j_2 <...<j_m}^N V_{j_1...j_m}^{(m)}
\end{align}

An immediate  way to obtain the generalized Hartree equation as the mean field limit is to let $N$ go to infinity and write the $N$ particle density as a tensor product of identical 1-particle densities. There is only one term for each $m$ in the BBGKY hierarchy equations  dominating this limit. For $W^{(m)}$ it is the term in the hierarchy equation corresponding to $l=m-1$. One gets the generalized Hartree equation from this crude procedure. 

Now we proceed to the BBGKY hierarchy equation for the full Hamiltonian $H_N$  in integral form 

\begin{align}
\gamma_N^{(k)}(t)= \mathcal{U}^{(k)}(t) | \phi(0) \rangle \langle \phi(0) |^{\otimes k}-i \sum_{m=2}^M \sum_{l=0}^{m-1} \frac{(N-k)!}{(N-k-l)!} \frac{1}{N^{m-1}} \nonumber
\\ \sum_{1 \leq j_1 < ... <j_{m-l}}^k \int_{0}^{t} ds \mathcal{U}^{(k)}(t-s) \text{tr}_{[k+1, ..., k+l]} [V_{j_1...j_m, k+1, ..., k+l}, \gamma_N^{(k+l)}(s)]
\end{align}
where $\mathcal{U}^{(k)}(t) \gamma_N^{(k)}=e^{-i \sum_{j+1}^k A_j t} \gamma_N^{(k)}(0) e^{i \sum_{j+1}^k A_j t}$ and the Hartree equation for k-particle density is

\begin{align}
i \partial_t | \phi(t) \rangle \langle \phi(t) |^{\otimes k} = \sum_{m=2}^M \sum_{j=1}^k \text{tr}_{[k+1, k+m-1]} [V^{(m)}_{j, k+1, ..., k+m-1} , | \phi(t) \rangle \langle \phi(t) |^{\otimes k+m-1}]
\end{align}
with the integral form

\begin{flalign}
&|\phi(t) \rangle \langle \phi(t) |^{\otimes k} =  \mathcal{U}^{(k)}(t) |\phi(0) \rangle \langle \phi(0) |^{\otimes k} \nonumber &
\\ &-i \sum_{m=2}^M \sum_{j=1}^k \int_0^{t}  \mathcal{U}^{(k)}(t-s) \text{tr}_{[k+1, k+m-1]} [V^{(m)}_{j, k+1, ..., k+m-1} , | \phi(s) \rangle \langle \phi(s) |^{\otimes k+m-1}]&
\end{flalign}

We are now ready to give the proof of main theorem stated in the first section. 

\noindent
\textbf{Proof of Theorem 1.} We can bound the projection of an arbitrary $k$-particle observable $J^{(k)}$ on the difference between BBGKY which is exact and the Hartree evolution.

\begin{flalign}
& \text{tr} J^{(k)}(\gamma_N^{(k)}(t)-|\phi(t) \rangle \langle \phi(t)|^{\otimes k}) \ \nonumber & \\
& = -i  \sum_{m=2}^M \sum_{l=0}^{m-2} \frac{(N-k)!}{(N-k-l)!)} \frac{1}{N^{m-1}} \sum_{1 \leq j_1 <...<j_{m-l}}^{k} \int_0^t ds \text{tr} (\mathcal{U}^{(k)}(t-s)J^{(k)} [ V^{(m)}_{j_1...j_{m-l}, k+1, ..., k+l}, \gamma_N^{(k+l)}(s)])  \nonumber &
\\ & -i  \sum_{m=2}^M  (\frac{(N-k)!}{(N-k-(m-1))!)}-N^{m-1}) \frac{1}{N^{m-1}} \sum_{j=1}^k \int_0^t ds \text{tr} (\mathcal{U}^{(k)}(t-s)J^{(k)} [ V^{(m)}_{j,k+1, ..., k+m-1}, \gamma_N^{(k+m-1)}(s)])  \nonumber  &
\\ &-i \sum_{m=2}^M \sum_{j=1}^k \int_0^t ds \text{tr} (\mathcal{U}^{(k)}(t-s)J^{(k)} [ V^{(m)}_{j,k+1, ..., k+m-1}, \gamma_N^{(k+m-1)}(s) - | \phi(s) \rangle \langle \phi(s)|^{\otimes (k+m-1)}])   \nonumber  &
\\ & =  h_1 + h_2 + h_3 
\end{flalign}

A useful fact we will frequently use is 

\begin{equation}
| \text{tr} (A [B,C]) | \leq 2|A| |B| \text{tr} |C|
\end{equation}

From now on set $k=1$, so that we are only interested in the single particle Hartree equation. The proof would require modifications otherwise. We will first bound the following term

\begin{align}
h_1=-i  \sum_{m=2}^M \sum_{l=0}^{m-2} \frac{(N-k)!}{(N-k-l)!)} \frac{1}{N^{m-1}} \sum_{1 \leq j_1 <...<j_{m-l}}^{k} \int_0^t ds \text{tr} (\mathcal{U}^{(k)}(t-s)J^{(k)} [ V^{(m)}_{j_1...j_{m-l}, k+1, ..., k+l}, \gamma_N^{(k+l)}(s)])
\end{align}

It follows from

\begin{align}
\text{tr} (\mathcal{U}^{(k)}(t-s)J^{(k)} [ V^{(m)}_{j_1...j_{m-l}, k+1, ..., k+l}, \gamma_N^{(k+l)}(s)])  \nonumber 
\\
\leq 2 |\mathcal{U}^{(k)}(t-s) J^{(k)}| |V^{(m)}| tr| \gamma_N^{(k+l)}(s)|=2 |J^{(k)}| |V^{(m)}|
\end{align} 
that

\begin{flalign}
 & | \sum_{m=2}^M \sum_{l=0}^{m-2} \frac{(N-k)!}{(N-k-l)!)} \frac{1}{N^{m-1}} \sum_{1 \leq j_1 <...<j_{m-l}}^{k} \int_0^t ds \text{tr} (\mathcal{U}^{(k)}(t-s)J^{(k)} [ V^{(m)}_{j_1...j_{m-l}, k+1, ..., k+l}, \gamma_N^{(k+l)}(s)]) |  \nonumber &
 \\ & \leq \sum_{m=2}^M \sum_{l=0}^{m-2} \frac{(N-k)!}{(N-k-l)!)} \frac{1}{N^{m-1}} \sum_{1 \leq j_1 <...<j_{m-l}}^{k} 2 |J^{(k)}| |V^{(m)}| t &
\end{flalign}

Setting $k=1$, this term has no contribution. Now, the bound for the second term 

\begin{equation}
h_2= -i  \sum_{m=2}^M  (\frac{(N-k)!}{(N-k-(m-1))!)}-N^{m-1}) \frac{1}{N^{m-1}} \sum_{j=1}^k \int_0^t ds \text{tr} (\mathcal{U}^{(k)}(t-s)J^{(k)} [ V^{(m)}_{j,k+1, ..., k+m-1}, \gamma_N^{(k+m-1)}(s)])
\end{equation}
is given by

\begin{flalign}
& | -i  \sum_{m=2}^M  (\frac{(N-1)!}{(N-m)!)}-N^{m-1}) \frac{1}{N^{m-1}} \sum_{j=1}^k \int_0^t ds \text{tr} (\mathcal{U}^{(1)}(t-s)J^{(1)} [ V^{(m)}_{1...m} \gamma_N^{(m)}(s)]) |   \nonumber &
\\ &\leq  \sum_{m=2}^M (1- \frac{(N-1)!}{(N-m)! N^{m-1}}) 2 |J^{(1)}| |V^{(m)}| t  \leq \sum_{m=2}^M \frac{m^2}{N} 2 |J^{(1)}| |V^{(m)}| t = \frac{2 |J^{(1)}|t}{N} (\sum l^2 \cdot V)&
\end{flalign}

Finally we want to bound the third term 

\begin{flalign}
& h_3=-i \sum_{m=2}^M  \int_0^t ds \text{tr} (\mathcal{U}^{(1)}(t-s)J^{(1)} [ V^{(m)}_{1, ...,m}, \gamma_N^{m}(s) - | \phi(s) \rangle \langle \phi(s)|^{\otimes m}]) & 
\end{flalign}

To proceed we need the following lemma, which can be proven by induction:

\begin{multline} 
  \gamma_N^{(m+1)}(t) - | \phi(t) \rangle \langle \phi(t)|^{\otimes (m+1)} =\sum_{l=1}^m (\gamma_N^{(l+1)}(t) -\gamma_N^{(l)}(t) \otimes \gamma_N^{(1)}(t) ) \otimes | \phi(t) \rangle \langle \phi(t)|^{\otimes m-l} 
 \\   +\sum_{l=0}^m \gamma_N^{(l)} \otimes (\gamma_N^{(1)} (t)- | \phi(t) \rangle \langle \phi(t)|) \otimes | \phi(t) \rangle \langle \phi(t)|^{\otimes m-l} 
\end{multline}

Inserting the lemma above we obtain

\begin{align}
& |-i \sum_{m=2}^M  \int_0^t ds \text{tr} (\mathcal{U}^{(1)}(t-s)J^{(1)} [ V^{(m)}_{1, ...,m}, \gamma_N^{m}(s) - | \phi(s) \rangle \langle \phi(s)|^{\otimes m}]) |  \nonumber & 
\\ & = | \sum_{m=2}^M \sum_{l=0}^{m-2} \int_0^t ds \text{tr}  (\mathcal{U}^{(1)}(t-s)J^{(1)} [ V^{(m)}_{1, ...,m},  \gamma_N^{(l)}(s) \otimes (\gamma_N^{(1)} (s)- | \phi(s) \rangle \langle \phi(s)|) \otimes | \phi(s) \rangle \langle \phi(s)|^{\otimes m-l-1}]|   \nonumber &
\\ & + | \sum_{m=2}^M \sum_{l=1}^{m-1} \int_0^t ds \text{tr}  (\mathcal{U}^{(1)}(t-s)J^{(1)} [ V^{(m)}_{1, ...,m}, (\gamma_N^{(l+1)}(s) -\gamma_N^{(l)}(s) \otimes \gamma_N^{(1)}(s) ) \otimes | \phi(s) \rangle \langle \phi(s)|^{\otimes m-l-1} ] |  \nonumber &
\\ & \leq 2 |J^{(1)}| (\sum l \cdot V) \int_0^t ds tr|\gamma_N^{(1)}(s) - | \phi(s) \rangle \langle \phi(s)||  \nonumber &
\\ & + | \sum_{m=2}^M \sum_{l=1}^{m-1} \int_0^t ds \text{tr}  (\mathcal{U}^{(1)}(t-s)J^{(1)} [ V^{(m)}_{1, ...,m}, (\gamma_N^{(l+1)}(s) -\gamma_N^{(l)}(s) \otimes \gamma_N^{(1)}(s) ) \otimes | \phi(s) \rangle \langle \phi(s)|^{\otimes m-l-1} ] | & 
\end{align}

Up to now correlation bounds have not been used. We will employ them to bound the second term appearing in the last expression. Recall that 

\begin{equation}
V^{(m)}=\sum_{i_1...i_m} w_{i_1...i_m}E_{1, i_1} \otimes... \otimes E_{m, i_m}
\end{equation}
for orthonormal bases $\{E_{j,k}\}_k$.  Then

\begin{flalign}
&  | \sum_{m=2}^M \sum_{l=1}^{m-1} \int_0^t ds \text{tr}  (\mathcal{U}^{(1)}(t-s)J^{(1)} [ V^{(m)}_{1, ...,m}, (\gamma_N^{(l+1)}(s) -\gamma_N^{(l)}(s) \otimes \gamma_N^{(1)}(s) ) \otimes | \phi(s) \rangle \langle \phi(s)|^{\otimes m-l-1} ] |  \nonumber &
\\ & \leq \sum_{m=2}^M \sum_{l=1}^{m-1} \int_0^t ds |\text{tr}  (\mathcal{U}^{(1)}(t-s)J^{(1)} [ V^{(m)}_{1, ...,m}, (\gamma_N^{(l+1)}(s) -\gamma_N^{(l)}(s) \otimes \gamma_N^{(1)}(s) ) \otimes | \phi(s) \rangle \langle \phi(s)|^{\otimes m-l-1} ] |  \nonumber &
\\ & \leq \sum_{m=2}^M \sum_{l=1}^{m-1} \sum_{i_1...i_m} |w_{i_1...i_m}|  \int_0^t ds |\text{tr}  (\mathcal{U}^{(1)}(t-s)J^{(1)}  \nonumber  &
\\ & [ E_{1,i_1} \otimes... \otimes E_{m,i_m} , (\gamma_N^{(l+1)}(s) -\gamma_N^{(l)}(s) \otimes \gamma_N^{(1)}(s) ) \otimes | \phi(s) \rangle \langle \phi(s)|^{\otimes m-l-1} ] |  \nonumber &
\\ & \leq 2 \sum_{m=2}^M \sum_{l=1}^{m-1} \sum_{i_1...i_m} |w_{i_1...i_m}|  \int_0^t ds |\text{tr}  (\mathcal{U}^{(1)}(t-s)J^{(1)}  E_{1,i_1} \otimes... \otimes E_{l+1,i_{l+1}}  (\gamma_N^{(l+1)}(s) -\gamma_N^{(l)}(s) \otimes \gamma_N^{(1)}(s) )  \nonumber &
\\ & \text{tr} (  E_{l+2,i_{l+2}} \otimes... \otimes E_{m,i_m} | \phi(s) \rangle \langle \phi(s)|^{\otimes m-l-1}) |  \nonumber & 
\\ & \leq 2 \sum_{m=2}^M \sum_{l=1}^{m-1} \sum_{i_1...i_m} |w_{i_1...i_m}|  \int_0^t ds |\text{tr}  (\mathcal{U}^{(1)}(t-s)J^{(1)}  E_{1,i_1} \otimes... \otimes E_{l+1,i_{l+1}}  (\gamma_N^{(l+1)}(s) -\gamma_N^{(l)}(s) \otimes \gamma_N^{(1)}(s) ) | &
\end{flalign}

Now, we use Corollary 2 to obtain

\begin{flalign}
&  | \sum_{m=2}^M \sum_{l=1}^{m-1} \int_0^t ds \text{tr}  (\mathcal{U}^{(1)}(t-s)J^{(1)} [ V^{(m)}_{1, ...,m}, (\gamma_N^{(l+1)}(s) -\gamma_N^{(l)}(s) \otimes \gamma_N^{(1)}(s) ) \otimes | \phi(s) \rangle \langle \phi(s)|^{\otimes m-l-1} ] |  \nonumber &
\\ & \leq  \sum_{m=2}^M \sum_{l=1}^{m-1} |\tilde{V}| \int_0^t ds \frac{32 l |J^{(1)}|}{N} (e^{4 (\sum l \cdot V)s} - 1 ) \leq M^3 \frac{8 | \tilde{V}| |J^{(1)}|}{ (\sum l \cdot V) N} (e^{4 (\sum l \cdot V)t} - 1 )&
\end{flalign}

Putting all these bounds together, we get

\begin{flalign}
&| \text{tr} J^{(1)}(\gamma_N^{(1)}(t)-|\phi(t) \rangle \langle \phi(t)||  \nonumber &
 \\ & \leq ( \sum l \cdot V) \int_0^t ds 2 |J^{(1)}| \text{tr}|\gamma_N^{(1)}- | \phi(s) \rangle \langle \phi(s) || + M^3 \frac{8 | \tilde{V}| |J^{(1)}|}{ (\sum l \cdot V) N} (e^{4 (\sum l \cdot V)t} - 1 ) + ( \sum l^2 \cdot V) \frac{2 | J^{(1)}| t}{N}  \nonumber &
 \\ & \leq ( \sum l \cdot V) \int_0^t ds 2 |J^{(1)}| \text{tr}|\gamma_N^{(1)}- | \phi(s) \rangle \langle \phi(s) || + M^3\lambda_V \frac{|J^{(1)}|}{2N} (e^{4 (\sum l \cdot V)t} - 1 )&
\end{flalign}

We can employ the operator definition of trace norm to get rid of $J^{(1)}$. 

\begin{align}
\text{tr}|A|=\text{sup}_{|J|=1} |\text{tr}JA|
\end{align}

Then

\begin{flalign}
&\text{tr}|\gamma_N^{(1)}(t)-|\phi(t) \rangle \langle \phi(t)|| \leq 2 ( \sum l \cdot V) \int_0^t ds  \text{tr}|\gamma_N^{(1)}- | \phi(s) \rangle \langle \phi(s) || + M^3\lambda_V \frac{1}{2N} (e^{4 (\sum l \cdot V)t} - 1 )&
\end{flalign}

Iterating $n$ times and letting $n$ go to infinity, we get the anticipated result:

\begin{align}
\text{tr}|\gamma_N^{(1)}(t)-|\phi(t) \rangle \langle \phi(t)|| \leq \frac{M^3}{N} \lambda_V  (e^{4 (\sum l \cdot V) t} - 1 )
\end{align}

\noindent
\textbf{Q.E.D.}



%
%

\bibliographystyle{unsrt}
\bibliography{Mean_field_references}

\end{document}